\documentstyle[prl,aps,amsfonts,amssymb,epsfig,floats]{revtex}
\begin{document}
\draft
\preprint{}
%
%following two lines are for two column format
\twocolumn[\hsize\textwidth\columnwidth\hsize\csname
@twocolumnfalse\endcsname

\title{ARPES study of Pb doped $Bi_{2}Sr_{2}CaCu_{2}O_{8}$ - a new Fermi surface picture}
\author{P. V. Bogdanov$^1$, A. Lanzara$^1{}^,{}^2$,  X. J. Zhou$^1$, S. A.
Kellar$^1$, D. L. Feng$^1$, E. D. Lu$^2$, H. Eisaki$^1$, J. -I. Shimoyama$^3$, K. Kishio$^3$, Z. Hussain$^2$, and Z. X. Shen$^1$
}
\address{
$^1$Department of Physics, Applied Physics and Stanford
Synchrotron Radiation Laboratory,\\ Stanford University, Stanford,
CA 94305, USA }
\address{
$^2$Advanced Light Source, Lawrence Berkeley National Lab,
Berkeley, CA 94720 }
\address{
$^3$Department of Applied Chemistry, University of Tokyo, Tokyo,
113-8656, Japan}

\date{\today}

\maketitle

\begin{abstract}
High resolution angle resolved photoemission data from Pb doped
$Bi_{2}Sr_{2}CaCu_{2}O_{8}$ (Bi2212) with suppressed superstructure is presented.
Improved resolution and very high momentum space sampling at
various photon energies reveal the presence of two Fermi surface
pieces. One has the hole-like topology, while the other one has its van Hove
singularity very close to $(\pi,0)$, its topology
at some photon energies resembles the electron-like piece. This result provides a unifying picture
of the Fermi surface in the Bi2212 compound and reconciles the
conflicting reports.
\end{abstract}
\vskip1pc]
\narrowtext

The Fermi surface plays an important role in understanding the
physics of any material. Among other things its shape and size
determine the type and number of charge carriers in the material
as well as the charge and spin dynamics. For example, in the context of
the Fermi liquid approach to high temperature superconductors
(HTSCs), Fermi surface topology is related to commensurate or
incommensurate nature of the neutron data \cite{NormanMode}.
Furthermore, detailed knowledge of the Fermi surface is essential
to determine the superconducting gap size and symmetry in the
superconducting state.

Angle resolved photoemission spectroscopy (ARPES)
is a unique tool to probe the Fermi surface of the HTSCs.
Over the last decade the HTSC system most extensively investigated by ARPES is
Bi2212 \cite{Olson,Dessau,Aebi,Ma,Loeser,White,Ding,Norman,Saini}.
However, the existence of superstructure in the $BiO$ layer and shadow
bands has made the Fermi
surface determination in this compound complicated, especially
around the $M (\pi,0)$ point, where main bands, superstructure
bands, and shadow bands cross the Fermi level. For several years
there was a general agreement for a
hole-like Fermi surface centered around $Y(\pi,\pi)$, mainly based on
ARPES experiments performed at $22 eV$ photon energy
\cite{Aebi,Ma,Loeser,White,Ding,Norman,Saini}. This hole-like Fermi surface picture was believed to
apply over the entire doping range studied by ARPES (from underdoped
samples with $Tc\approx15K$ to overdoped samples with
$Tc\approx68K$). Later a vigorous discussion started,
with experiments utilizing $33 eV$
photon energy suggesting an electron-like Fermi
surface centered around the $\Gamma$ point
\cite{Chuang,Feng,Gromko}. Other groups disputed electron-like
Fermi surface reports, dismissing the observed
results by invoking the interplay of matrix element
effects with $BiO$ layer superstructure and shadowbands
\cite{Fretwell,Mesot,Borisenko,Legner}. Finally, recent
reports on Bi2212 using 22 eV photons demonstrated
the presence of the two Fermi surfaces in the
material due to bonding and antibonding interaction of CuO planes
\cite{FengBS,ChuangBS}. It is of great importance to reconcile all the reported results and to
resolve the uncertainty in the Fermi surface of Bi2212 - the compound
extensively studied and the source of many significant results.

Because the main discrepancy originates in the $(\pi,0)$ region,
where the superstructure effect of the $BiO$ layer is strongest, a
definitive resolution of the Fermi surface issue can be found by
studying the Pb-doped Bi2212 system. In this compound Pb
is doped into the $BiO$ plane disrupting the $BiO$ plane
modulation and removing the superstructure complication near the $(\pi,0)$
region \cite{Borisenko,Legner,Schwaller}.

In this Letter we present results of Fermi surface mapping of
Pb-doped Bi2212 with high energy resolution and very high $k$
space sampling. We used various photon energies and different
methods to determine the Fermi surface. Our 22 eV data complement
recent reports on the existence of two Fermi surface pieces.
The photon energy dependence reveals that the relative intensity
depends strongly on photon energy. While the bonding Fermi surface
has a clear hole-like topology, the antibonding piece has its van
Hove singularity very close to $(\pi,0)$ and its Fermi surface is
electron-like as seen at some photon energies. This
result contradicts earlier reports of a single
universal Fermi surface in this compound
\cite{Ding,Borisenko,Legner}. On the other hand our data provides
a unifying foundation for understanding the controversies about
the Fermi surface of this important superconductor, as different
reports stress different aspects of the global Fermi surface
features.

\begin{figure*}[t!]
 \centerline{\epsfig{file=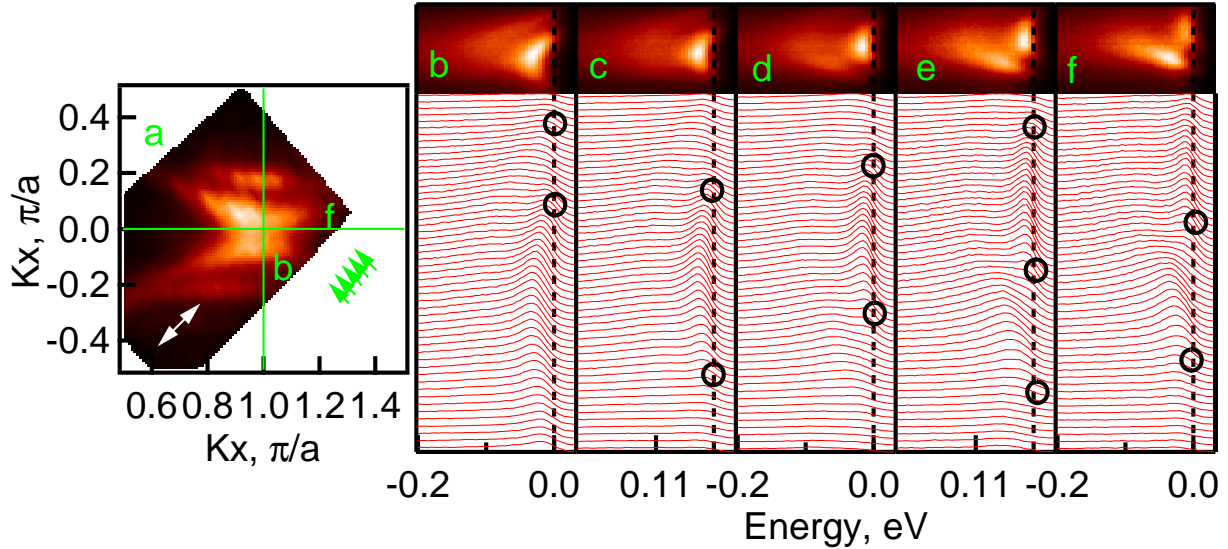,width=7.03in,clip=}}
 \vspace{0in}
 \caption{Panel a) shows the spectral intensity at the Fermi level
 for data collected with 22 eV photons. White arrow shows the
 light polarization. Panels b) - f) show EDCs
 along the cuts indicated in panel a). EDCs are
 stacked top to bottom in the direction of the arrows.}
 \label{22eV}
\end{figure*}

ARPES data have been recorded at beamline $10.0.1.1$ of the
Advanced Light Source utilizing $55, 44, 33, 27$ and $22$ $eV$ photon energy
in $4\cdot10^-{}^1{}^1$ $Torr$ vacuum. The sample was kept in the fixed position relative to the beam
polarization, and the analyzer was rotated. The beam polarization
was in the sample plane perpendicular to $\Gamma-Y$ direction, with
beam nearly at grazing incidence with the sample surface. We
used a Scienta SES 200 analyzer in the angle mode, where cuts parallel
to $\Gamma-Y$ direction are carried out. The momentum resolution was $\pm0.06
\AA^{-1}$ in the scan direction and $\pm0.19 \AA^{-1}$ in the
perpendicular direction for 55 eV photon energy and better for other energies,
and the energy resolution was $7-18$ $meV$. An
extensive and fine sampling mesh with more than 4000 EDCs for
each photon energy was collected. The slightly overdoped Pb-doped Bi2212
($Tc = 84K$) and overdoped Pb-doped Bi2212 ($Tc = 70K$)
were grown using the floating-zone method. The single crystalline
samples were oriented by using Laue diffraction ex situ
and cleaved in situ in vacuum. The samples were measured at 100K
(Fig. 1,2 and 4) and 20K (Fig. 3). The Fermi energy
was obtained from the EDCs of polycrystalline $Au$.

In panel a) of Fig. 1 we show the map of spectral intensity
at Fermi energy ($E_F$) obtained at 22 eV photon energy.
The white arrow shows the polarization of radiation with
respect to the crystal surface. To
determine the spectral intensity map we divide each EDC by the
integrated signal intensity from a 100 meV window above the
$E_F$, which comes from higher order synchrotron
light and is proportional to photon flux. The
normalized EDCs represent electron spectral function weighted by
the Fermi function and matrix element\cite{Hufner}. Highest
intensity points in the spectral intensity map at the Fermi energy
give one method for determining Fermi surface.  In panels
1b)-1f) we plot raw EDC data obtained along select cuts shown in
panel a) by green arrows. Here Fermi surface crossing is
defined as the location in the momentum space where the
intensity of the spectral feature decreases drastically and
the leading edge crosses the Fermi level. From both the near $E_F$ spectral weight image
plot and from the EDCs one clearly sees two Fermi surfaces, as
indicated by solid ovals in panels b to f. Concomitant presence of two Fermi
surface pieces is consistent with recent data recorded at a
similar photon energy \cite{FengBS,ChuangBS}

\begin{figure}[b!]
 \centerline{\epsfig{file=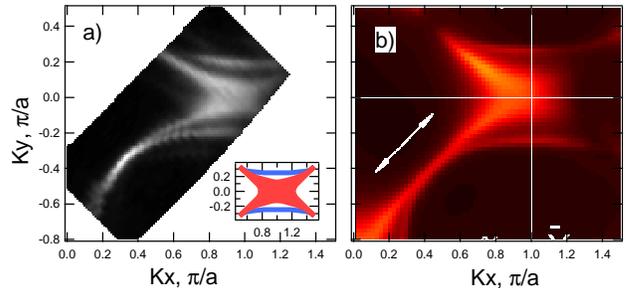,width=3.4in, clip=}}
 \vspace{0in}
 \caption{Panel a)shows experimental spectral intensity map at 12
 meV BE collected using 22 eV photons.  Panel b) shows calculated
 ARPES intensity for Bi2212 in the same experimental conditions by
 Bansil $\it et. al$. Light polarization is given
 by the arrow in this panel and is the same as in panel a).
 Inset in panel a) schematically indicates
 two Fermi surface pieces.}
 \label{Comparison}
\end{figure}
\begin{figure*}[t!]
 \centerline{\epsfig{file=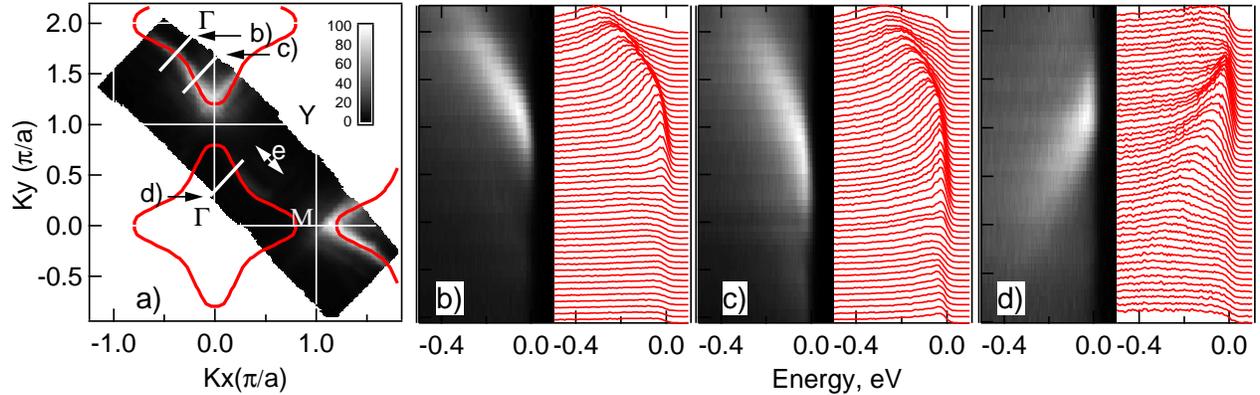,width=7.03in,clip=}}
 \vspace{0in}
 \caption{Gray scale image in panel a)
 shows the spectral intensity at the Fermi level collected with 55
 ev photons. The white arrow shows the polarization of radiation with
respect to the crystal surface. Arrows and thick lines indicate the cuts presented in
 panels b) - d). Red line indicates the Fermi surface shape.
 Panels b) - d) show data along select cuts in the Brillouin zone.
 Spectra are stacked top to bottom parallel to $(\pi,\pi)-(0,0)$ direction.
 Left side of each panel shows 2D plot with brightness proportional
 to signal intensity, while right side shows
 corresponding EDCs equally spaced in vertical direction for
 clarity.
 }
 \label{55eV}
\end{figure*}
Fig. 2a) shows 22 eV data from another sample taken in a more
extended k-space area. This data were taken at 20K in the
superconducting state, where the spectral weight around M point is
suppressed at the Fermi level due to the sc gap opening, so the
map shown corresponds to 12 meV BE. This map can effectively be
used to indicate the underlying Fermi surface. Although the
maximum gap is larger than the energy window, finite resolution
still reveals the underlying Fermi surface, and larger integration
window does not change the picture. This data, as well as the
normal state data in panel a) of Fig. 1, show striking resemblance
to theoretical simulation by Bansil et al. \cite{Bansil} for the
same photon energy and polarization, as shown in panel b). The
simulation uses first-principles one-step photoemission model
calculation and comes up with two bands for two adjacent $CuO_2$ planes in
a unit cell. These two bonding and antibonding bands
give rise to two Fermi surfaces. The
bonding piece is an outer hole-like piece,
indicated by blue lines in the inset of panel a). On the other
hand the antibonding piece, indicated by the red area in the inset
of panel a), is hard to judge from these data, because the saddle
point of the band is very close to $E_F$ at $(\pi,0)$
\cite{BansilTalk}. The simulation also indicates that the image plot is
very similar whether this piece of the Fermi surface is hole-like
or electron-like, i.e. whether the Van Hove saddle point is above
or below $E_F$. The absence of superstructure complication
in our data and the striking similarity between experiment and
theory strongly suggest that there are indeed two pieces of the
Fermi surface in Pb Bi2212.

The Fermi surface seen under other measurement conditions turns out
to be very different. Fig. 3 shows data recorded at 55 eV under the same
measurement geometry. Panel a) shows the map of spectral intensity
at $E_F$ in the momentum space. In panels b), c)
and d) we plot raw data obtained along the select
cuts shown in panel a) by thick white lines. Cut b) is close to
the nodal direction in the second Brillouin zone, while cuts c) and d) are
cuts equidistant from $M$ point. The sampling density in the cuts is very
high and is representative of the sampling density of the entire k-space studied.
The high quality data clearly shows a quasiparticle dispersing towards the
Fermi level, eventually crossing it and disappearing. While the
intensity map in panel a) hardly shows any features in the first
zone, panel d) clearly shows a well-resolved feature crossing the
Fermi level, similar to panel c), with the overall intensity a
factor of 10 lower than that in panel c). In fact, all features
seen in the second zone are observed in the first zone as well.
The Fermi surface shape emerging for this photon energy is
electron-like. We have  confirmed this result with data taken in all
three Brillouin zones by using three complimentary methods:
intensity map at $E_F$, the traditional method of tracking the
EDCs, and the sharpest drop in $n(\vec k)$.
The contrast in data from Figures 1-3 immediately suggests that
the 55 eV data picks out the inner piece of the Fermi surface.

To investigate the photon energy dependence of the FS further,
we collected data at other photon
energies and in Fig. 4 we plot the measured spectral intensity maps at $E_f$. In panel
a) we plot our spectral intensity map
collected at 27 eV. We see an electron-like FS in the first zone, and the
spectral weight in the second zone is significantly suppressed.
This complements the 55 eV data in panel d). In panel b)
at 33 eV photon energy spectral intensity map shows strong suppression of the spectral
weight at the M point and can be interpreted as either hole-like
and electron-like. This data is quite different from
earlier results recorded at another geometry
\cite{Saini,Chuang,Feng}.
44 eV data in panel c) looks very similar to 22
eV data in panel a) with bilayer split Fermi surfaces. Spectral
weight map FS results are supported by individual EDC analysis.

There's a very unusual photon energy related variation in
the ARPES data. With the large unit cell size in the normal
direction ($c\approx30.6\AA$), one would expect periodicity with
1-2 eV steps in photon energy for the energy range studied.
Surprisingly, this is not the case. Existence of bilayer split
Fermi surface means strong interaction between the two $CuO_2$ planes
in Bi2212 unit cell. It is reasonable to assume that the variation
in the data with the photon energy is also driven by the separation
of the layers $d\approx3.4\AA$, with corresponding
$Kz =\frac{2\pi}{d}=1.85\AA$. This would be more consistent with
the observed variations with large photon energy intervals.
 Calculations similar to
\cite{Bansil} for a range of photon energies
will be very useful to understand this phenomena.

\begin{figure}[b!]
 \centerline{\epsfig{file=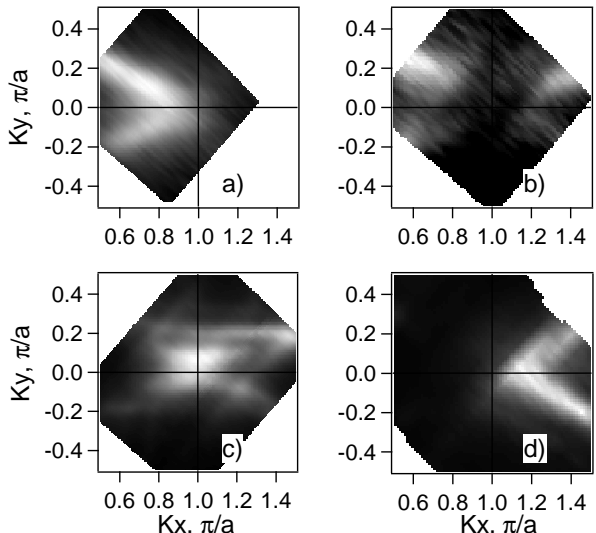,width=3.4in, clip=}}
 \vspace{0in}
 \caption{In this figure spectral intensity maps at the Fermi
 level for different photon energies are shown. Panel a) shows
 data collected at $27 eV$, panel b) shows $33 eV$ data, panel
 c) shows $44 eV$ data and panel d) $55 eV$ data.}
 \label{allFS}
\end{figure}

Our results, in particular the variation of the Fermi surface picture
with photon energies and the observation of the bilayer splitting,
contradict earlier reports from Pb doped Bi2212. Previous data \cite{Borisenko,Legner}
from the same material at different photon energies were interpreted as an
evidence for the universal hole-like Fermi surface.  We attribute
this discrepancy mostly to poorer energy (70 meV compared to ours 7-18
meV) and momentum ($0.094\AA^{-1}$ compared to ours $0.006\times 0.019\AA^{-1}$)
resolution used in that study for photon energies other then 22.4
eV. Attempts were also made to support a single universal hole-like Fermi surface
picture by invoking matrix element arguments\cite{Fretwell,Mesot}.
While a band calculation did
show a strong spectral intensity variation with photon energy \cite{Bansil},
it is not applicable to Fermi surface determination at a
particular photon energy. What really matters is the matrix
element variation in a small region in the same Brillouin zone,
which is actually small \cite{Bansil}. The unambiguous evidence
for two Fermi surface pieces show the matrix element argument for
single universal hole-like Fermi surface to be misleading and incorrect.

The polarization setup used in our experiment
is favorable for observing the bilayer splitting \cite{Bansil}.
Observation of the splitting
of the bands due to the interaction
between the layers was a long standing problem. Earlier
photoemission results were interpreted as evidence for the
absence of the bilayer splitting \cite{Ding}. Our data in Fig. 1,
2 and 4, collected with much better energy and momentum
resolution, show the splitting to be present, even for the samples
with Tc not far from optimal.

The picture emerging from the above is clear: there are two pieces
of the Fermi surface, one of them clearly hole-like. The other piece is
different, as it lies very close to the special point where Fermi
surface changes from hole-like to electron-like with small change
in chemical potential or $\vec K_Z$. This is the underlying reason for this
piece to behave slightly differently at different photon energies.
This picture provides a unifying foundation for all the
controversial reports on the FS shape in Bi2212.
Bi2212 Fermi surface has always been attributed to the $CuO_2$ plane, and because
doping with Pb does not change the $CuO_2$ plane, the Fermi
surface in Pb-doped and Pb-free compounds should be the same. Our finding indicates that the
accepted picture of a single hole-like Fermi surface in Bi2212 for the entire
doping range studied is incorrect. Instead, there exists another
Fermi surface piece that is at the boundary between hole and
electron character.

We would like to thank J. D. Denlinger for the help with data
analysis software. The experiment was performed at the Advanced
Light Source of Lawrence Berkeley National Laboratory, supported
by DOE' Office of Basic Energy Science, Division of
Materials Science with contract DE-AC03-76SF00098. The
Stanford work was supported by NSF grant through the Stanford
MRSEC grant and NSF grant DMR-9705210. The SSRL's
work was also supported by the Office's Division of Materials
Science.


\begin{references}
\bibitem{NormanMode} M. R. Norman, preprint, cond-mat / 9912203
\bibitem{Olson} C. G. Olson {\sl et al.}, Science {\bf 245}, 731 (1989)
\bibitem{Dessau}D. Dessau {\sl et al.}, Phys. Rev. Lett. {\bf 71}, 2781
(1993)
\bibitem{Aebi} P. Aebi {\it et al.}, Phys. Rev. Lett. {\bf 72} 2757 (1994)
\bibitem{Ma} Jian Ma {\it et al.}, Phys. Rev. B {\bf 51}, 3832 (1995)
\bibitem{Loeser} A. G. Loeser {\it et al.}, Science {\bf 273}, 325 (1996)
\bibitem{White} P. J. White {\it et al.}, Phys. Rev. B {\bf 54}, 15669 (1996)
\bibitem{Ding} H. Ding {\it et al.}, Phys. Rev. Lett. {\bf 76}, 1533 (1996)
\bibitem{Norman} M. R. Norman {\it et al.}, Nature {\bf 392}, 157 (1998)
\bibitem{Saini} N. L. Saini {\it et al.}, Phys. Rev. Lett. {\bf 79}, 3467 (1997)
\bibitem{Chuang} Y. -D. Chuang {\it et al.}, Phys. Rev. Lett. {\bf 83}, 3717 (1999).
\bibitem{Feng} D. L. Feng {\it et al.}, preprint, cond-mat / 9908056.
\bibitem{Gromko} A. D. Gromko {\it et al.}, preprint, cond-mat / 0003017.
\bibitem{Fretwell} H. M. Fretwell {\it et al.}, Phys. Rev. Lett. {\bf 84}, 4449 (2000)
\bibitem{Mesot} J. Mesot {\it et al.}, preprint, cond-mat / 9910430.
\bibitem{Borisenko} S. V. Borisenko {\it et al.}, Phys. Rev. Lett. {\bf 84}, 4453 (2000)
\bibitem{Legner} S. Legner {\it et al.}, preprint, cond-mat / 0002302.
\bibitem{FengBS} D. L. Feng {\it et al.}, cond-mat/0102385
\bibitem{ChuangBS}Y. -D. Chuang {\it et al.}, cond-mat/0102386
\bibitem{Schwaller} P. Schwaller {\it et al.}, J. Elec. Spec. Rel. Phen., {\bf
76} 127 (1995).
\bibitem{Hufner} S. Hufner, {\it Photoemission Spectroscopy}, Springer-Verlag, New York, (1995)
\bibitem{Bansil} A. Bansil {\it et al.}, Phys. Rev.
Lett. {\bf 83} 5154 (1999).
\bibitem{BansilTalk} A. Bansil {\it et al.}, private communication

\end{references}
\end{document}